# Subband Random Sensing Grant Free Uplink for URLLC in Unlicensed Spectrum[*]

Yonghong Zeng, Yuhong Wang, Sumei Sun, and Yugang Ma

Institute for Infocomm Research, A*STAR, Singapore

Emails: {yhzeng, wangyh, sunsm, mayg}@i2r.a-star.edu.sg

*Abstract*— In this paper, we propose a novel scheme called subband random sensing (SRS) grant free uplink for ultra-reliable low-latency communication (URLLC) in unlicensed spectrum. The SRS grant free uplink creatively combines the subband sensing, user grouping and random access, which allows a user sensing the unlicensed spectrum to use available sub-resources in a wideband and reduce collisions with other users. The new scheme overcomes the severe spectrum wastage problem of the semi-persistent scheduling (SPS) uplink adopted by new radio (NR) release 16 in applications with sporadic traffic. Compared with the contention based grant free uplink, the new scheme achieves much lower collision probability, which directly leads to higher reliability. Analysis and simulations are provided to prove the exceptional performances.

## I. INTRODUCTION

Ultra-reliable low-latency communication (URLLC) is one of the key components in 5G [1-2]. URLLC is particularly important for industrial internet of things (IIOT) that includes widespread applications such as [3-5]:

- Industrial process automation, factory automation (for example, real time control of robots);
- Automation of smart grid energy distribution, detection of and restoration from faults;
- Automated intelligent transport systems (for example, real-time coordination among autonomous vehicles).

On the other hand, due to the proliferation of the wireless communications, radio spectrum becomes increasingly scarcer in recent years. Thus, shared usage of unlicensed spectrum has become an effective way to maximize the spectral efficiency. In general, spectrum sharing has the following advantages: (1) reduce cost as unlicensed spectrum is free; (2) allow for new deployment scenarios; (3) increase spectrum utilization. However, for any user using the unlicensed spectrum, there is no guarantee of immediate access to the spectrum as the spectrum is shared. In addition, the transmission in unlicensed spectrum may be interfered by signals from other users sharing the same spectrum. As a result, the communication latency and reliability may become unpredictable. How to realize URLLC in unlicensed spectrum is especially challenging and needs to be addressed [6-8].

In recent years, URLLC has been a hot research topic in the communication community [1-5]. In the latest 3GPP 5G new radio (NR) standard Rel-16 [2], a few technologies have been proposed to support the URLLC. These include: (1) large subcarrier spacing (SCS) that reduces the transmission time interval (TTI); (2) mini slots that reduces the transmission and alignment time; (3) preemption that gives priority packet guaranteed slot for transmission; and (4) grant free uplink that allows uplink traffic to transmit without grant request. However, almost all research works and standards on URLLC have assumed to use licensed spectrum [2-5]. The special difficulties and challenges of URLLC by using unlicensed spectrum have barely been touched.

It is well known that ultra-low uplink latency is one of the most difficult tasks in URLLC [9-13]. Traditionally an uplink user needs to request the base station (gNB in 5G) for the grant of transmission before emitting its signal. This "grant based" uplink in general cannot meet the delay requirement due to the time spent in the grant process [9, 10]. In recent years, "grant free" uplink schemes have been proposed to by-pass the grant process and allow an uplink user transmitting without a "grant" [9-13]. However, such grant free uplink schemes are originally proposed for licensed spectrum [11]. They do not provide any mechanism to sense the spectrum and handle the spectrum disruption when using unlicensed spectrum. Thus, their performances are not guaranteed when using unlicensed spectrum.

In this paper, we first briefly review the existing grant free uplink schemes and their problems. Then, we propose a novel uplink scheme for URLLC called subband random sensing (SRS) grant free uplink. The SRS grant free uplink intriguingly combines the subband sensing, user grouping and random access, which allows a user to sense the spectrum, use available sub-resources in a wide band and reduces collisions with other users. It is shown that the new scheme overcomes the spectrum wastage problem of the *semi-persistent*

---

[*]This work is supported by RIE2020 ADVANCED MANUFACTURING AND ENGINEERING (AME) INDUSTRY ALIGNMENT FUND-PRE-POSITIONING (IAF-PP) under grant number A20F8a0044.

*scheduling* (SPS) uplink adopted in the NR Rel-16 and achieves much lower collision probability than the *contention based* grant free uplink in applications with sporadic traffic. Lower collision probability directly leads to higher reliability.

The remaining part of the paper is organized as follows. The special challenges for URLLC in unlicensed spectrum is briefly reviewed in Section II. In Section III, the SRS grant free uplink is proposed. The performance evaluation is given in Section IV. Finally, conclusions and future researches are given in Section V.

## II. SPECIAL CHALLENGES FOR URLLC IN UNLICENSED SPECTRUM

In this section, we briefly review the special challenges for URLLC in unlicensed spectrum, which include non-guaranteed spectrum, the cost due to the listen before talk (LBT) and the time delay in the process of request for grant.

### A. Non-guaranteed transmission opportunity and reliability

Fig. 1 shows an example when the spectrum is shared by the NR in unlicensed band (NR-U) [2] and WiFi. Both of them use LBT to share the same channel. When the WiFi is using the channel, the NR-U has to defer its transmission until the WiFi has finished the transmission. If NR-U happens to have a critical message to send, it must wait and the waiting time depends on how fast the WiFi finishes its transmission. Let $T_w$ be the WiFi packet length. NR-U must wait at least $T_w$ time to start its transmission. Within the WiFi network, there may be other users also competing for the channel on the moment. Thus, it is not guaranteed that NR-U will get the channel even after the $T_w$ time. In addition, it is hard to guarantee the reliability as there may be abrupt interference from transmissions of others due to the LBT error or hidden node problem.

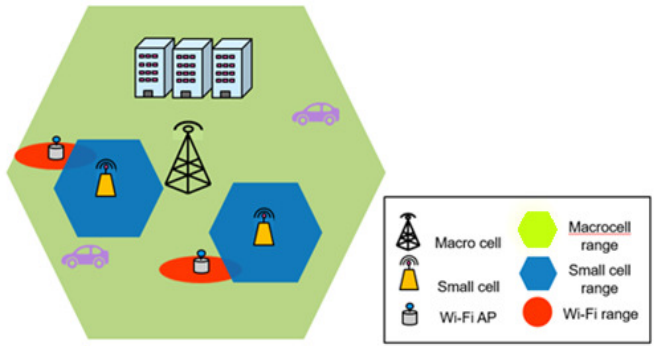

Fig. 1 Spectrum sharing of NR-U/WiFi in unlicensed band

### B. Additional delay due to the LBT

LBT is the first step for any system to use the unlicensed spectrum [2, 14-16]. Obviously, if the time spent for LBT is already longer than one millisecond (1ms) (URLLC delay limit specified in NR Rel-16), there is no way to achieve the delay requirement of URLLC. In [8, 15, 16], the time spent for LBT in current standards has been studied in detail. It is shown that the time for LBT is directly related to the probability of channel availability, $p$. Higher $p$ means lower latency. On the other hand, $p$ is inversely proportional to the traffic load of co-channel users. If there are other systems that use the same spectrum, their traffic load will be the deciding factor in the LBT process. While it is possible to adjust the traffic of our own system, it is impossible to control the traffic of other systems. It is shown in [8] that $p$ must be higher than 0.95 to guarantee less than 1ms latency for LBT. That is, the channel must be sensed free with no lower than 95% probability in the LBT process. This is a very stringent requirement in practice. Basically, it says that there is almost no co-channel user from other systems. Thus, we can say that it is virtually impossible to realize URLLC by using just a single channel with the conventional LBT scheme. Thus, it is necessary to use multiple transmit/receive chains [17, 18] or a wideband with subband sensing to achieve URLLC.

### C. Even more challenging for URLLC uplink

It is well known that the ultra-low latency for uplink is even more challenging [9-13]. For uplink transmissions in unlicensed spectrum, LBT is also required before transmission in most cases in current standards [2, 14], but different LBT schemes may be used at different situations. For example, if the uplink transmission is within the base station acquired Maximum Channel Occupancy Time (MCOT), the user equipment (UE) can perform a category-2 (CAT-2) LBT that is a fixed 25 microsecond (us) duration [14, 2]. Furthermore, in the MulteFire specifications [14], a UE can bypass the LBT and transmit immediately if the transmission of uplink control information starts no later than 16us after the end of the downlink transmission. However, in other situations, UE needs to perform category-4 (CAT-4) LBT before transmission.

Other than the LBT, any uplink transmission needs coordination/cooperation of the gNB. Conventionally an uplink user needs to request the gNB for the grant of transmission before transmitting its signal. The process is like this: a UE sends a transmission request to gNB, the gNB receives the request and checks the available resources for the UE uplink, and the gNB informs the UE the allocated resources or "no/deferring transmission" decision. The UE then can transmit at the allocated resources (the given time and frequency). Such "grant based" uplink in general cannot meet the URLLC delay requirement due to the extra delay in the grant process [9, 10].

## III. SRS GRANT FREE UPLINK

As discussed in the above section, we see that the LBT and the grant process are the two additional challenges to achieve URLLC uplink in unlicensed spectrum. In recent years, there have been researches on "grant free" uplink, which try to by-pass the grant process and allow an uplink user to transmit without a "grant" [9-13, 19]. In this section, we will first briefly review the existing grant free uplink schemes and their problems. Then we will propose a new scheme and compare it with the existing ones.

*A. Exisiting grant free uplink and their problems*

Although there are various grant free uplink schemes, the SPS uplink [2, 11, 12] and contention based grant free uplink [9, 11-13] are the most popular ones.

**SPS grant free uplink**. SPS uplink is a grant free uplink that pre-schedules radio resources for uplink [2, 11, 12]. Any uplink user can only transmit in the given time and frequency whenever it has data to transmit. If the user's uplink transmission time is *pre-determined* in advance, SPS scheme is quite efficient. However, let us consider more challenging applications where the uplink traffic is unpredictable, which is common in IIOT and called sporadic traffic [9]. The SPS scheme can be *very inefficient for sporadic traffic*. In fact, for using SPS uplink in sporadic traffic, to meet the latency requirement, the time gap $T_g$ between two allocated resources must be below the maximum allowed uplink delay $T_u$, $T_g \leq T_u$. Assume that the size of a URLLC packet is $P_u$ bits. Let $N$ be the number of URLLC users associated with the gNB. Each user must transmit its URLLC data within $T_u$ time whenever it has data to transmit. Thus, within the $T_u$ duration, any URLLC user must be allocated a resource for transmission.

In practice, except the URLLC, there may be other applications sharing the same spectrum. Thus we assume that URLLC can only use part of the communication time, say, $T_{au}$. That is, all URLLC users must have a chance to transmit their data within the $T_{au}$ duration. Furthermore, in general UE needs to do LBT before transmission. Let $T_{LBT}$ be the average time spending on the LBT. Thus, the total time spent on LBT for all the UEs is $NT_{LBT}$, if the resource allocation for different users is based on time division multiple access (TDMA). Hence, the actual time for uplink transmission is only $T_{au} - NT_{LBT}$. Thus the resources (data rate) allocated for uplink must be at least $\frac{NP_u}{T_{au} - NT_{LBT}}$ bits per second (bps), even if the channel is always available after LBT. For example, let $P_u = 32$ byte, $T_u = 0.5$ ms (based on the 1ms delay requirement for both uplink and downlink), $T_{au} = 0.5T_u$ (half of the time is used for URLLC), and $T_{LBT} = 25$ us (CAT-2 LBT specification). The supported number of users and required data rate are shown in Fig. 2, where only 9 users can be supported with data rate at least 92.16 Mbps.

Problems of SPS grant free uplink are summarized as follows.

- There may be no enough resources to meet the requirement in URLLC when the number of users are large, as the gap between two scheduled resources $T_g$ must be smaller than the maximum allowed uplink delay.
- For sporadic traffic, the UE may not have data to transmit at the pre-scheduled resources, thus resources are wasted.

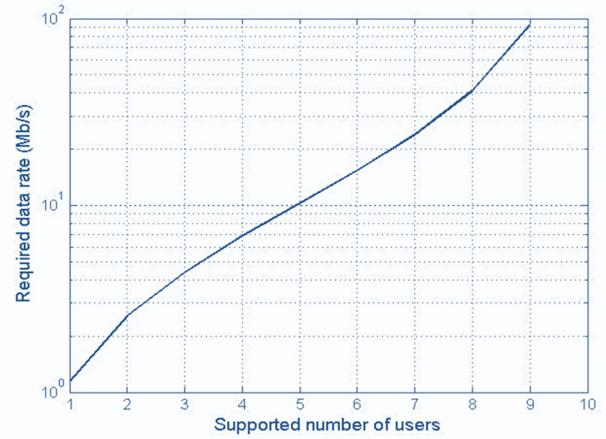

Fig. 2 Required data rate for SPS uplink (TDMA based resource allocation)

**Contention based grant free uplink**. Contention based grant free uplink was proposed in [9, 11-13] to overcome the resource wastage problem in SPS uplink. It is a modification and generalization of the SPS uplink. The major difference to the SPS uplink is that *a block of resources is pre-scheduled for a user group (a group of users)*. Any user in the group randomly chooses a resource in the pre-scheduled shared resources for uplink. Problems of contention based grant free uplink are summarized as follows.

- There is no collision avoidance mechanism, as the users in one group transmit randomly. The collision probability is high [11, 12].
- In [11, 19], multi-user detection combined with non-orthogonal multiple access (NOMA) and advanced hybrid automatic repeat request (HARQ) are used to resolve the collision problem. However, this brings additional overheads and high complexity.

Both the SPS and contention based grant free uplink are originally proposed for *licensed band* [11]. When used for *unlicensed band* [12], the channel may be unavailable at anytime, which prevents the UEs to transmit at any of the resources in the scheduled resources and induces additional delays.

*B. SRS grant free uplink*

We group the UEs such that UEs in the same user group (UG) can hear from each other. This can be done based on the geolocation information of the users. An example of grouping is shown in Fig. 3, where the 11 users in the cell are divided into 3 groups. The grouping may need to change dynamically based on the latest location/statistics information. After grouping, we allocate different (orthogonal) resources for different UGs. As the users in different groups use orthogonal time/frequency resources for transmissions, their signals will not interfere with each other.

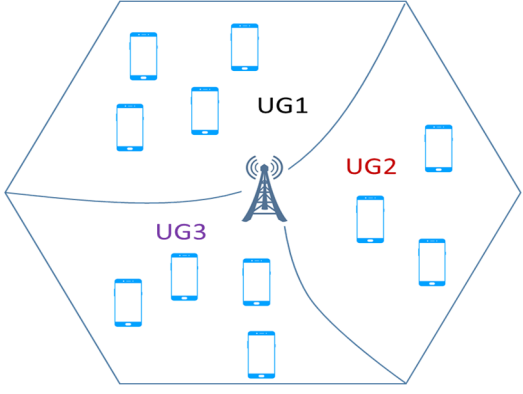

Fig. 3 User grouping based on geolocation information

The resources in a UG is further divided into $K$ orthogonal sub-resources (SRs), where each SR is just enough for uplink of one UE's data plus LBT, as shown in Fig. 4.

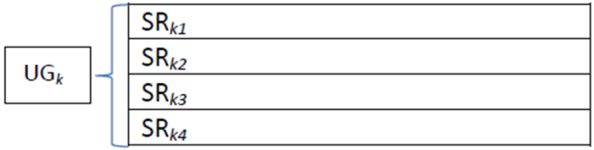

Fig. 4 Resource division within a user group

Every UE must sense the SRs allocated for its UG before transmission (subband sensing). It randomly chooses a sensing duration $T_i + T_a l$, where $T_i$ and $T_a$ are the initial and additional sensing durations respectively, $l$ is a random integer number in $[0, L-1]$, and $L$ is the number of sensing slots. This subband random sensing is shown in Fig. 5. After sensing, it randomly chooses a free SR to transmit its uplink data including the identity. If there is no free SR, it defers the transmission. After choosing the free SR, the UE may need to defer sometime to synchronize its OFDM symbols. In order to occupy the channel and allow other UEs sensing its transmission, the UE can transmit a short flag signal immediately after sensing. Finally, advanced HARQ could be used for error control and retransmission to further guarantee the reliability [11, 19].

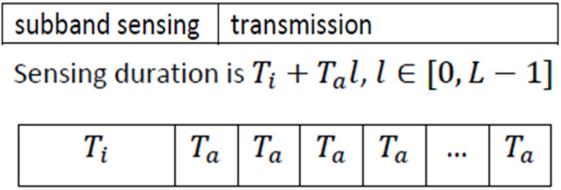

Fig. 5 The subband random sensing scheme

*C. Analysis of the scheme*

Let $N_u$ be the number of UGs, $N_g$ be the number of UEs in the same UG and $K$ be the number of SRs. Note that a SR is the minimal resource required to support one user's uplink. In general, $N_g$ can be much larger than $K$ for sporadic traffic. SPS uplink can only support $N_u K$ users, while the proposed method can support $N_u N_g$ users, with $N_g \gg K$. However, there are possibilities of collision (two users choose the same SR for transmission). The probabilities of collision is related to $N_g$, $K$, sensing procedure, and traffic arrival model.

When two or more UEs in the same UG transmitting at the same time with the same SR, there is a collision. The proposed subband sensing has a collision avoidance mechanism: a UE can sense the signals of other UEs in the same UG and then avoid transmission at the same SR. If two UEs choose different sensing durations, the UE that senses longer will detect the presence of UE that senses shorter. The transmissions of two UEs have a collision if and only if they choose the same SR and the same sensing duration. Note that when a collision happens, the gNB cannot decode the signals, which causes retransmissions with large delay.

If a UE senses all the $K$ SRs and they are occupied, it will not have chance to transmit (here and after, we also treat this as a collision). This happens when the number of active users at the moment is larger than $K$ (include active users associated to the gNB and active users from other systems). **Note:** If the UE has sensed all the channels and found no free channel, it can continue to sense if the allowed maximum sensing time is not passed. This is to grasp the *chance of newly released channel* by other coexistence systems (WiFi, NR, ….).

When there are $n$ UEs that want to transmit in the same UG and at the same time slot, the collision probability $P_c(n)$ of UE's transmissions is

$$P_c(n) = 0, \; n = 1 \qquad (1)$$

$$P_c(n) = \left(1 - \frac{(K-1)\cdots(K-n+1)}{K^{n-1}}\right) \cdot \left(1 - \frac{(L-1)\cdots(L-n+1)}{L^{n-1}}\right), n = 2, \ldots, K \qquad (2)$$

$$P_c(n) = 1, \; n > K \qquad (3)$$

Obviously, if no user transmitting in the UG, the probability of collision is 0, that is, $P_c(0) = 0$. Let $P_u(n)$ be the probability of $n$ active UEs in the same UG (**note:** $P_u(n)$ is related to traffic model). Then the overall collision probability is

$$P_{oc} = \sum_{n=0}^{N_g} P_u(n) P_c(n) \qquad (4)$$

where $N_g$ is the maximum number of UEs in a UG, and $\sum_{n=0}^{N} P_u(n) = 1$.

IV. PERFORMANCE EVALUATION

The proposed SRS grant free uplink intriguingly combines the subband sensing, user grouping and random access, which allows UE to use available sub-resources in a large band and reduces collisions among different users.

Compared with SPS uplink, the advantages/disadvantages of the proposed method are as follows.

- SPS uplink can only support $N_u K$ users, while the proposed method can support $N_u N_g$ users, with $N_g \gg K$.
- When the SPS uplink is used in unlicensed band, the scheduled resource may not be available due to the contention of the channel by other systems (for example, WiFi, NR-U from a different operator, Zigbee, …). The proposed method partially solves the problem by subband sensing and resource sharing.
- However, there are possibilities of collision (two users choose the same SR for transmission) in the proposed SRS uplink. The probabilities of collision is related to $N_g$, $K$, sensing procedure, and traffic arrival model.

Compared with the contention based grant free uplink, the advantages of the proposed method are as follows.

- The SRS uplink and contention based uplink support the same number of users, but the SRS uplink has a collision avoidance mechanism through the subband random sensing, which reduces collision probability considerably (shown later).
- When the contention based uplink is used in unlicensed band [12], the scheduled shared resource may be unavailable due to the contention of the channel by other systems. The proposed method partially solves the problem by subband sensing

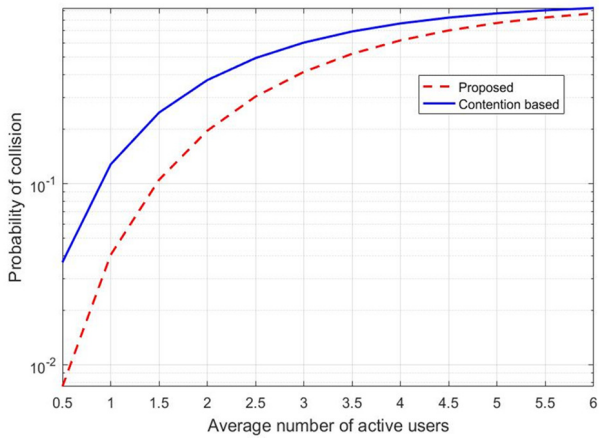

Fig. 6 Probability of collision versus average number of active UEs within a group ($K = 3$, $L = 9$, $N_g = 15$)

Here we show some simulation results on the collision probability for the proposed method and the contention based method. The simulation condition is as follows. (1) Each UG has $N_g$ UEs; (2) There are $K$ SRs in a UG; (3) The number of URLLC active users follows Poisson distribution with parameter $\lambda$, where $\lambda$ is the average number of active users. We use Matlab to randomly generate the active users and its packets. Fig. 6 to Fig. 8 show the probability of collision versus the average number of active users within a group for three different cases: (1) $K = 3$, $L = 9$, $N_g = 15$; (2) $K = 5$, $L = 11$, $N_g = 20$, (3) $K = 8$, $L = 15$, $N_g = 40$. Obviously, for all the considered cases, *the proposed method reduces collision probability considerably. Lower collision probability directly contributes to higher reliability (lower BER).*

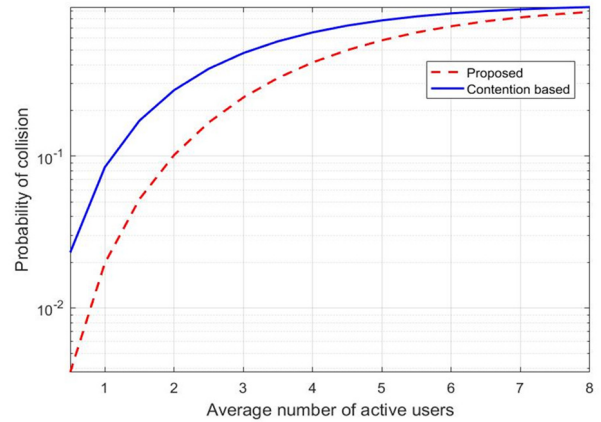

Fig. 7 Probability of collision versus average number of active UEs within a group ($K = 5$, $L = 11$, $N_g = 20$)

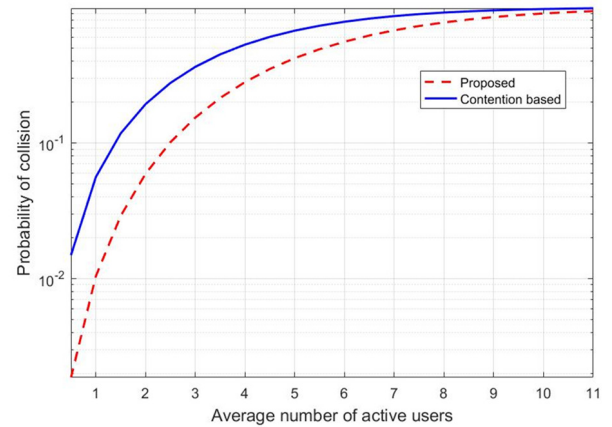

Fig. 8 Probability of collision versus average number of active UEs within a group ($K = 8$, $L = 15$, $N_g = 40$)

## V. CONCLUSION AND FUTURE RESEARCH

In this paper, we have proposed a brand new uplink scheme called SRS grant free uplink for URLLC. The SRS grant free uplink creatively combines the subband sensing, user grouping and random access, which allows a user sensing the unlicensed spectrum, using available sub-resources in a wideband and reducing collisions with other users. It is shown that the new scheme overcomes the severe spectrum wastage problem in the SPS uplink adopted in NR Rel-16 and achieves much lower collision probability compared with the contention based grant free uplink in applications with sporadic traffic. Analysis and simulations are provided to prove the superior performances.

As part of our future research, we will conduct more analysis and comparison based on realistic environment for these schemes. We will also study the effective methods for user grouping and analyze the implementation cost.

# REFERENCES

[1] 5G Americas white paper, "New services & applications with 5G ultra-reliable low latency communications", Nov. 2018.

[2] LTE release 16, https://www.3gpp.org/release-16, 3GPP, July 2020.

[3] Qualcomm, "Ultra-reliable and low latency for industrai automation", https://www.qualcomm.com/media/documents, 2018.

[4] A. Aijaz, and M. Sooriyabandara, "The Tactile Internet for Industries: A Review", Proceedings of the IEEE, Vol. 107, No. 2, pp. 414-435, Feb. 2019.

[5] A. Nasrallah et al., "Ultra-Low Latency (ULL) Networks: The IEEE TSN and IETF DetNet Standards and Related 5G ULL Research", IEEE Comm. Surv. Tutorials, 21(1): 88-145. 2019.

[6] G. J. Sutton et al., "Enabling Ultra-Reliable and Low-Latency Communications through Unlicensed Spectrum", IEEE Network, vol. 32, no. 2, pp. 70 -77, Mar/April 2018.

[7] Y. H. Wang , Y. H. Zeng , S. M. Sun , and Y. Nan. "High Accuracy Uplink Timing Synchronization for 5G NR in Unlicensed Spectrum", IEEE Wireless Comm. Letters (Early Access ), Nov. 2020.

[8] Y. H. Zeng, Y. H. Wang, S. M. Sun and K. Yang, "Feasibility of URLLC in Unlicensed Spectrum", IEEE Asia Pacific Wireless Communications Symposium, Aug. 2019.

[9] S. Ali, N. Rajatheva, and W. Saad, "Fast Uplink Grant for Machine Type Communications: Challenges and Opportunities", IEEE Comm. Magazine, vol.57, no. 3, pp. 97 – 103. Mar. 2019.

[10] T. Jacobsen et al., "System Level Analysis of Uplink Grant-Free Transmission for URLLC", IEEE Globecom Workshops (GC Wkshps), Dec. 2017.

[11] C. Wang, Y. Chen, Y. Wu and L. Zhang, "Performance Evaluation of Grant-free Transmission for Uplink URLLC Services", IEEE 85th Vehicular Technology Conference (VTC Spring), 2017.

[12] R. Karaki, A. Mukherjee, and J.-F. Cheng, "Performance of Autonomous Uplink Transmissions in Unlicensed Spectrum LTE", IEEE Globecom, Dec. 2017.

[13] N. H. Mahmood, R. Abreu, R. Bohnke, M. Schubert, G. Berardinelli, and T. H. Jacobsen, "Uplink Grant-Free Access Solutions for URLLC services in 5G New Radio", 16th International Symposium on Wireless Communication Systems (ISWCS), pp. 607-612, Oulu, Finland, 2019.

[14] MulteFire Alliance, "Multefire release 1.0 technical paper: A new way to wireless," Tech. Rep.: https://www.multefire.org/

[15] R. M. Cuevas et al., "On the Impact of Listen-Before-Talk on Ultra-Reliable Low-Latency Communications", IEEE GlobeCom, UAE, Dec. 2018.

[16] G. J. Sutton , R. P. Liu , and Y. J. Guo, "Delay and Reliability of Load-Based Listen-Before-Talk in LAA", IEEE Access, vol. 6, pp. 6171-6182, March 2018.

[17] J. J. Nielsen, R. Liu, and P. Popovski, "Ultra-Reliable Low Latency Communication (URLLC) using Interface Diversity", IEEE Transactions on Comm., vol. 66, no.3, pp.1322-1334, March 2018.

[18] C. She et al., "Improving Network Availability of Ultra-Reliable and Low-Latency Communications with Multi-Connectivity", IEEE Trans. on Communications, vol. 66, no. 11, pp. 5482 – 5496, Nov. 2018.

[19] R. Kotaba, C. N. Manchon, T. Balercia, and P. Popovski, "How URLLC can Benefit from NOMA-based Retransmissions", IEEE Trans. Wireless Comm., vol. 20, no. 3, pp. 1684-1699, March 2021.